**Observation of anisotropic fragmentation in methane subjected to femtosecond radiation**

J. Strohaber, F. Zhu, A. A. Kolomenskii, and H. A. Schuessler

*Texas A&M University, Department of Physics and Astronomy, College Station, TX 77843-4242, USA*

*\*Corresponding author: jstroha1@gmail.com*

We present experimental results on the ionization/dissociation of methane in femtosecond pulses of radiation. Angular and intensity dependent yields of singly and doubly charged species were measured using an imaging mass spectrometer. The measured data shows that all fragments yields exhibit some degree of anisotropy as a result of them being preferably ejected parallel to the polarization direction. Additionally, an anomalous perpendicular fragmentation pattern is found for $CH_2^{2+}$. We find evidence of multiple dissociation mechanisms including statistical decay, field assisted dissociation and Coulomb explosion.

The interactions of molecular systems with radiation are interesting and rich with phenomena that are not found in atomic systems. Such effects include charge resonant enhanced ionization (CREI) [1], dissociation ionization (DI) and Coulomb explosion (CE) [2], and angular-dependent effects from aligned or oriented molecules [3]. As experiments have shown, the history of the interaction of methane with various forms of radiation has been fascinating and replete with perplexing photophysical processes. Methane is the simplest hydrocarbon and is the primary building block for more complex organic molecules, thereby making its photodynamical properties of fundamental interest to a wide spectrum of disciplines. It is the main ingredient of natural gas, and it is a substantial contributor of the Greenhouse effect.

Methane has been investigated under a number of experimental conditions, and during its interaction with radiation it is possible to produce the parent molecular ions and their dissociation products $CH_n^{m+}$ ($n = 0-4$; m=1,2) and $H_n^{m+}$ ($n = 1,2$; m=0,1) [4]. Of these ions, theoretical and experimental results have demonstrated that the ions $CH_3^{2+}$ and $CH^{2+}$ are unstable and do not appear in time-of-flight (TOF) spectra. When using nanosecond pulses of radiation, experiments revealed the complete absence of the parent molecular ion $CH_4^+$ in the TOF spectrum [5]. Ionization and binding energies of $CH_4$ ($CH_4^+$) are 12.6 eV (19.7 eV) and 4.3 eV (1.2 eV), and because the photon energy in these experiments was 2.3 eV, $CH_4$ tends to dissociate into $CH_3^+$ and H. Upon ionization with picosecond pulses [532 nm (2.3 eV)] [6], the parent ion $CH_4^+$ appeared in the spectrum; however, this time fragmental $CH_3^+$ was missing. This phenomenon was attributed to charge migration facilitated by the laser field resulting in the instability of $CH_3^+$. The most striking results were obtained in experiments with few-cycle femtosecond pulses [7]. For an intensity of $\sim 10^{14}$ W/cm$^2$ and pulse duration of 110 fs, all ions appeared in the spectrum; however, when using 8 fs pulses at the same peak intensity, all ions were missing from the spectrum except for $CH_4^+$ and $CH_3^+$. Predicted dissociation probabilities from field-assisted dissociation (FAD) indicate that 8 fs is insufficient time for $CH_4^+$ to dissociate. The $CH_3^+$ yields were attributed to electron recollision with $CH_4^+$. The methane ion $CH_4^{2+}$ was absent because for short pulses the molecules do not have enough time to stretch to a critical internuclear separation distance for enhanced ionization of $CH_4^+$ to occur [7]. Other experiments have used: (i) chirped pulses, which showed enhanced fragmentation [8]; and (ii) relativistic intensities, which showed atomic-like ionization behavior [9]. While previous experiments have provided a relatively complete and well documented foundation describing the ionization/dissociation of methane, important details may still be missing and currently no adequate theoretical description has been reported.

In this paper we present angular and intensity dependent results for the ionization/dissociation of methane that show new features. To obtain a more precise picture of the underlying processes, we used an imaging TOF mass spectrometer, capable of avoiding spatial averaging effects that would have otherwise concealed the details presented in this work. To the best of our knowledge, the observed effects have not been reported experimentally or theoretically. The conclusions drawn in this work are based on experimental results supplemented by molecular optimization and potential energy curve (PEC) calculations with Gaussian 09 at the level of RHF/URH and 6-311G/cc-pVTZ.

Interpreting the interactions of methane with femtosecond radiation is not a straightforward task for a number of reasons. A natural question that arises in regard to dissociation concerns how charge distributes amongst the dissociation products. In other words, which fragment gets the electron? According to the FAD model [10], this question is addressed by investigating distortions of PECs when the electric field is directed from atom X to atom Y or vice versa. In the case of methane, the primary dissociation channel $CH_4^+ \rightarrow CH_3^{(1-m)+} + H^{m+}$ ($m = 0,1$) in the FAD description is predicted to produce the products with $m = 1$. This results from the fact that distortions of the PEC along a C—H bond, when the laser field is pointing in the direction of the H atom, are greater than for the opposing case. Thus, the field facilitates charge migration towards the central C atom. In other work [4], it was suggested that the primary dissociation channel was the case with $m = 0$, i.e., fragmentation with neutral H atoms. Neutral hydrogen production can be concluded by comparing the ionization energies of the neutral fragments of $CH_n$ (9.8, 10.4, 10.6, 11.2) eV [11] to that of atomic H (13.6 eV), which shows a preference for the binding electron attachment to $H^+$. Short vibrational periods of methane, which are on the order of 15 fs [12], must also be considered. One consequence of this short vibrational time period is that the internuclear separation distances can vary during the interaction with femtosecond radiation, and the validity of the frozen nuclei approximation must be questioned. Such questions were raised in the case of charge resonance enhanced ionization of $H_2^+$ [13]. It has been put forth that the lack of observation of the elusive double peak structure in the ionization of $H_2^+$ is partially due to the failure of the Born-Oppenheimer approximation. To obtain an adequate description of the ionization/dissociation dynamics of methane in femtosecond radiation, more extensive experimental data are needed.

In our experiments, femtosecond radiation having a wavelength of 800 nm, pulse energy of 1 mJ and duration of 50 fs was focused into an ionization chamber with a background pressure of $5 \times 10^{-9}$ mbar. For angular scans, an achromatic half-wave plate was placed at the entrance window to the chamber. This wave-plate was rotated using a motorized rotary stage having an angular resolution of 0.2 mrad (8401-M, New Focus). Methane gas was introduced into the chamber through a leak valve to an operating pressure of $10^{-7}$ mbar. The design and operation of our TOF setup is described in detail in Ref [14], and relevant discussions on volumetric weighting can be found in Refs [14—16]. In brief, this type of spectrometer was designed to measure intensity-resolved ionization yield by collecting ions from a micron-sized volume within the focus, over which the intensity is essentially constant. Ions were detected by a chevron-type multichannel plate (Photonis), and the output was sent to a multiscalar counter having a 100 ps timing resolution (MCS6, FAST ComTec). For intensity scans, the half-wave plate was positioned before the grating compressor of the chirped-pulse amplification laser to adjust the power of the output radiation. All experiments were fully automated by LabView. Because of possible degeneracies in the methane spectrum due to atmospheric gases $O_2$, $N_2$, and water, $^{12}CH_4$ and $^{13}CH_4$ were used to resolve potential ambiguities in the origin of the mass to charge peaks.

Figure 1 shows a typical time-of-flight spectrum of methane ionized with 50 fs laser pulses at a peak laser intensity of $6.3 \times 10^{14}$ W/cm$^2$. The data shows two main ionization stages: the first cluster consists of singly-charged ions $CH_n^+$ ($n = 0-4$), and the second cluster is that of the doubly-charged ions $CH_{2n}^{2+}$ ($n = 0, 1, 2$). The fragments $CH_{2n+1}^{2+}$ ($n = 0, 1$) are absent from the spectrum, and calculated PECs (Gaussian 09, UHF, 6-113-G) have confirmed low potential barrier heights of ~0.04 eV and ~0.03 eV, which make these fragments unstable [7]. Higher charge states $m > 2$ were not observed in the spectrum at the measured intensities. Intensity dependent ionization yields in Fig. 2(a) indicate that dissociation of $CH_n^+$ follows a step-wise process such as $CH_4^+ \rightarrow CH_3^+ + H$, then $CH_2^+ \rightarrow CH^+ + H$ and so on. Calibration of the intensity was made by measuring the saturation intensity of $Xe^+$, and the $1/e$ saturation intensity of $CH_4^+$ ($CH_4^{2+}$) was experimentally determined to be $\sim 1.4 \times 10^{14}$ W/cm$^2$ ($2.64 \times 10^{14}$ W/cm$^2$). Methane's ionization energies (IEs) and degenerate p-type molecular orbitals are similar to the IEs and atomic orbitals of xenon and motivate comparison of these yields with those calculated from atomic ionization theories such as ADK or PPT [17]. Adiabatic ionization energies of 12.6 eV, 19.7 eV [11] and 28.2 eV (This last value was extrapolated by comparing ionization energy of $CH_4$ with those of Xe) for $m = 1, 2, 3$ charge states were used in these calculations. The results of the calculations are shown in Fig. 2(b), and are in agreement with the measured yield curves. The theoretical probability curve of $CH_4^{2+}$ was multiplied by 0.04 for comparison with the measured data. Similar conclusions for $CH_4^+$ have been previously reported [18]. For the highest intensities, $CH_4^{2+}$ has reached saturation and is being depleted. This depletion is due to either dissociation of $CH_4^{2+}$ or triple ionization. Methane trications do not appear in the measured spectrum, and our attempts to optimize the molecular geometry of $CH_4^{3+}$ using Gaussian 09 failed at all levels of calculation suggesting that this ion is unstable.

Angular dependences of ion yields of methane have been previously measured [10, 19]. In these previous experiments no anisotropies were found in the methane fragments. Besides being able to predict the preference of one dissociation channel over another, the FAD model, by construct, results in fragments being ejected along the laser polarization direction as a result of pulling off chemical bonds that are aligned with the electric field vector [10]. Figure 3 shows angular dependent ion yields of $CH_n^+$ ($n=1-3$) measured at an intensity of $1.2 \times 10^{14}$ W/cm$^2$ and normalized to $CH_4^+$ to compensate for pressure effects. In this work, all angular dependent ion yields have been fitted to $r = r_0 + a\cos^{2n}(\theta) + b\sin^{2m}(\theta)$ (black curve), where $r_0$ (red curve) is the fraction of polarization-independent ions, $a$ (blue curve) is the fraction of ions being ejected along the polarization, $b$ (green curve) is the fraction of ions ejected perpendicular to the polarization, and $n$ and $m$ are positive integers. All yields in Fig. 3 are isotropic and show no indication of polarization dependence. The primary dissociation channel $CH_4^+ \rightarrow CH_3 + H^+$ predicted by FAD produces neutral $CH_3$ fragments, and conclusion must be based on measured anisotropies in $H^+$ [10]. However, for this intensity and level of statistics, $H^+$ ions were not observed in the angular scans. For $CH_n^+$ ($n=0,1,2$), the FAD model predicts that the fragments will be charged and in principle observable in the spectrum [10]. Since no anisotropies were observed, we conclude that FAD is not the dominant mechanism at this intensity level.

The data presented in Fig. 3 are, however, consistent with predictions based on the quasi-equilibrium theory (QET) and Rice–Ramsperger–Kassel–Marcus (RRKM) theory, which calculate dissociation rates based on available internal energy partitioned amongst the degrees of freedom of the molecules [4]. This statistical theory is polarization independent and results in an isotropic fragmentation pattern.

At an intensity of $3.6 \times 10^{14}$ W/cm$^2$, doubly-charged ions are produced. All fragment ion yields exhibit some degree of anisotropy (Fig. 4). These anisotropies in the fragmented ions are consistent with predictions made by FAD and exhibit dissociation along the polarization direction. Because of the presence of trace amounts of $H_2O$ in our experiments, ionization of water in atmospheric gas was also measured. Comparison of these results to those of the methane data showed that the amount of $H^+$ produced from the residual $H_2O$ was negligible. When comparing the yields of $H^+$ to those of $CH_n^+$, the $H^+$ yield was found to sit between $CH_2^+$ and $CH^+$ (Fig. 2), suggesting that the FAD mechanism is playing a role in dissociation. By normalizing the peak of the $CH_4^{2+}$ ion yield to that of $CH_4^+$ (Fig. 2), we found that the normalized $CH_4^{2+}$ ion yield crossed the parallel $H^+$ yield at $1.3 \times 10^{14}$ W/cm$^2$. Hydrogen yields below this intensity are greater than those of the normalized $CH_4^{2+}$, indicating that the $H^+$ ions below this intensity are originating from dissociation of singly-charged ions. These $H^+$ ions show a large anisotropy, and we conclude that FAD is taking place, but to a lesser degree than the more dominant QET mechanism. Above this intensity, double ionization of $CH_4$ becomes important, and anisotropies in $CH_n^+$ and $H^+$ will have a contribution from the Coulomb explosion of $CH_n^{2+}$.

By normalizing each fragment in Fig. 2 (parallel polarization) to $CH_4^+$ and doing the same for perpendicular polarization, comparison of relative ion yields between the different polarizations could be made. For $CH_3^+$ the two curves (parallel and perpendicular) have the same relative yields at lower intensities but diverge at an intensity of $\sim 1.3 \times 10^{14}$ W/cm$^2$. Above this intensity, $H^+$ production from $CH_4^{2+}$ is expected to occur. We conclude that the anisotropy in the $CH_3^+$ ion yields is due to Coulomb explosion $CH_4^{2+} \rightarrow CH_3^+ + H^+$. For $CH_n^+$ ($n=0,1,2$) production, the yields are separate at all measured intensities. The FAD model predicts that the preferred reaction channel is $CH_n^+ \rightarrow CH_{n-1}^+ + H$ ($n = 0-3$), a result consistent with our findings.

The largest anisotropy is found for $CH_2^{2+}$. This fragment must originate from a molecule that has more than 2 hydrogen atoms $CH_{n>2}^{m+}$ to account for kinetic energy release (KER). Therefore, $CH_2^{2+}$ cannot directly come from the double ionization of $CH_2^+$ produced in the statistical dissociation of $CH_3^+$. It is widely accepted that highly charged molecular ions dissociate due to Coulomb explosion [7, 10, 19]. We surmise that this is occurring in the dissociation of the ions $CH_n^{2+}$. For $CH_2^{2+}$, the most likely precursor candidates will be species having molecular geometries that facilitate entrance of fragmental $CH_2^{2+}$ into the TOF. The geometry of $CH_4^{2+}$ is nearly square planar and therefore requires the simultaneous removal of two adjacent H atoms. For this to occur, the polarization must bisect two $C_2$ symmetry axes. The geometry of $CH_3^{2+}$ is $C_{2v}$ with one bond elongated from 1.125 to 1.555 Angstrom (principle axis). This geometry allows for a C—H bond to be aligned with the polarization direction. We therefore attribute the $CH_2^{2+}$ anisotropy to Coulomb explosion of $CH_3^{2+}$, in which the product $CH_2^+$ can promptly ionize (IE=19.6 eV).

For our maximum obtainable laser intensity of $6.3 \times 10^{14}$ W/cm$^2$, the anisotropy of the fragment yields $CH_n^+$ have further increased due to Coulomb explosion of the ion $CH_n^{2+} \rightarrow CH_{n-1}^+ + H^+$ (Fig. 5). A new structure has appeared in the $CH_2^{2+}$ yield at this intensity.

This structure is a distribution resulting from $CH_2^{2+}$ fragments being ejected perpendicular to the laser polarization and is increasingly becoming the dominant contribution to the $CH_2^{2+}$ yield.

Similar perpendicular fragmentation has been observed in femtosecond experiments with $N_2O$ and $CS_2$ [20—22]; however, all these instances have involved *atomic* species being ejected perpendicular to the field. In each of these cases, perpendicular fragmentation was attributed to light-induced bending of the initially linear molecules to an angle of ~ 140 degree. Coulomb explosion then resulted in the center atom being ejected perpendicular to the laser field direction. To the best of our knowledge our results are the first reported instance of *molecular* fragments being ejected perpendicular to the laser field. As with parallel fragmentation, the perpendicular fragments must come from species with more than 2 hydrogen atoms $CH_{n>2}^{2+}$. Planar $CH_3^{m+}$ fragments can be ruled out, because Coulomb explosion is facilitated when one of the C—H bonds is along the polarization direction, and for $CH_3^{m+}$ no geometric configuration could be found to account for the perpendicular pattern. This means that $CH_4^{m+}$ is a likely candidate, but requires the simultaneous ejection of two hydrogen atoms. Geometric optimization of $CH_4^{2+}$ (UHF, cc-pTVZ) shows that the methane dication is planar, but two opposing hydrogen atoms are bent toward each other with a $H_{(1)}-C-H_{(3)}$ angle of ~153 degrees (here subscripts denote hydrogen position). For $CH_2^{2+}$ to be ejected perpendicular to the polarization, the electric field vector must be perpendicular to the linear $H_{(2)}-C-H_{(4)}$ part of $CH_4^{2+}$. We conclude that the precursor for the perpendicular fragmentation is $CH_4^{2+}$, and the mechanism is Coulomb explosion, followed by prompt ionization.

A one-to-one mapping of spatial positions to arrival times hold well for atoms in the absence of pressure effects; however, the spectrometer is a kinetic filter, and the inclusion of KER must be addressed to ascertain its influence on the arrival times of the fragments. From energy and momentum conservation, the kinetic energy of a di-fragmental dissociation is $KE_M[\text{eV}] = KER[\text{eV}]/(1+M/m)$. Here $M$ is the mass of the heavier fragment, and $m$ is that of the lighter fragment. As an example, if we let $KER = 1.2$ eV for the $CH_4^+ \rightarrow CH_3^+ + H$ dissociation channel of methane, than the $CH_3^+$ fragment acquires $KE = 0.07$ eV of kinetic energy and that of $H$ is $KE = 1.13$ eV. For ions ejected along the TOF axis, the difference in time-of-flight of the front and backward directed ions compared to those having zero initial kinetic energy can be approximated as

$$\Delta t_{\text{front}}^{\text{back}}[\text{ns}] = \alpha \frac{x_R}{\Phi_R} \sqrt{\frac{M}{Q}} \left( KE \frac{\partial \tau_H}{\partial x}\bigg|_{x=x_0} \pm \sqrt{2KE} \right). \tag{1}$$

Here $\alpha = 0.101805052$, $x_R[\mu\text{m}]$ and $\Phi_R[\text{eV}]$ are the position and voltage of the repeller plate, $\partial \tau_H / \partial x = 0.47$ ns/μm is the slope of the TOF dispersion curve for the hydrogen ion, and $KE[\text{eV}]$ is the kinetic energy of the fragment. An ion produced at position $x_0$ with kinetic energy $KE$ will enter the slit with the kinetic energy of an ion produced at the position $x_1 = x_0 + KE(x_R/\Phi_R)$. Using $\Delta t = \sqrt{M/Q}\left(\partial \tau_H / \partial x\right)\Delta x$ [14] we arrive at the first term in Eq. 1. The second term is the TOF difference (to enter the slit) between two identical ions having the same initial position $x_0$ with only one having a nonzero initial kinetic energy. The plus sign is for the backward directed ion and the negative is for the frontward directed ions. For comparison of TOFs between different charge states and masses, all ions are scaled to the position of H by $t_1^1 = t_M^Q \sqrt{Q/M}$, which removes the charge-to-mass ratio appearing in Eq. (1). For our parameters, Eq. (1) reduces to $\Delta t_{front}^{back} = 0.09 KE \pm 0.28\sqrt{KE}$. As an example, the release of the $KER = 4.3$ eV results in the following initial kinetic energies $KE_{CH_3} = 0.27$ eV, $KE_H = 4.03$ eV. According to Eq. (1) these kinetic energies correspond to time shifts of $\Delta t_{CH_3} = [\text{back, front}] = [0.2, -0.1]$ ns, and $\Delta t_H = [1.0, -0.2]$ ns. When the ions are ejected perpendicular to the TOF axis, the amount of kinetic energy that will allow a particle to still pass through the slit is $KE[\text{meV}] = 0.3 L^2$ (where $L$ is the slit width). For our setup, the slit is around 10 microns so $KE[\text{meV}] = 30$ meV. As a final note, the tradeoff of retaining spatial information is that the large extraction voltage appearing in the denominator in Eq. (1) results in smaller time shifts of front and backward directed fragments, which could otherwise have been used for kinetic energy measurements. However, previous kinetic energy experiments of this type have not revealed the reported anisotropies.

Images of ions produced along the $x$-direction in the focus are shown in Fig. 6. Panels (a—d) are yields of the singly-charged species $CH_n^+$ ($n = 0-4$), panels (e—h) are for the doubly-charged species $CH_{2n}^{2+}$ ($n = 0, 1, 2$), and panel (i) is for hydrogen. For comparison, all yields were made to overlap with each other. The overlap was achieved by fitting known ions in the spectrum to $t = \tau_H \sqrt{M/Q} + t_{off}$ and determining $\tau_H$ and $t_{off}$. Because the peak of the $H^+$ yield is expected to experience a noticeable shift due to KER, only the carbon-containing ions were used for calibration. Using this equation, the arrival times of the carbon-containing species of methane were scaled to overlap with the expected position of the hydrogen ion $\tau_H$, followed by subtracting $\tau_H$ to center the yields on zero. The images in panels (a—c, f) show a depletion for the higher intensities, which are around zero time. These scans were taken at our highest peak laser intensity, and no apparent distortions can be seen in the recorded positions of the yields with the exception of $H^+$. To investigate the amount of KER, all peaks in the spectrum were fitted to suitable functions to track and adjust their temporal locations. The result of this procedure is shown in Fig 7. The $CH_n^{m+}$ ions are found to overlap around $\tau_H \sim 4.913$ μm. In addition to an overall increase in the arrival time relative to the expected time $\tau_H$, $H^+$ and $H_2^+$ show an increasing temporal displacement when the polarization is parallel to the TOF axis. From these time shifts, energies can be computed for both parallel and perpendicular polarizations. For parallel polarization, a time shift of 4 ns was found for $H^+$ corresponding to $KE = 27$ eV, and for $H_2^+$ we find $KE = 4.4$ eV (1 ns). For perpendicular polarization the values are $KE = 39.7$ eV (5.5 ns) for $H^+$ and $KE = 18.9$ eV (3 ns) for $H_2^+$.

In conclusion, we have measured the ionization/dissociation products of methane subjected to 800 nm, 50 fs pulses of radiation. The parent molecular ion yields agree well with atomic-like ionization theories. It was found that all dissociation products eventually exhibit an anisotropic fragmentation pattern. For the singly-charged fragments, the dominant dissociation mechanism is consistent with statistical theories and to a lesser extent it depends on field-assisted dissociation and Coulomb explosion. Doubly charged ions of $CH_2^{2+}$ exhibit intensity dependent anisotropic yields with maxima both parallel and perpendicular to the polarization. The data presented here provides much needed information in obtaining a better understanding of the complex interaction of methane with femtosecond radiation. Our detection setup is capable of recording each individual event for each laser pulse, however, because we are not able to sufficiently distinguish between front and backward directed ions, our coincident analysis was not useful in determining the dissociation channels. As a final note, we measured similar data for $CH_{4-n}Cl_n$ ($n = 0-4$) and did not observe anisotropic fragmentation patterns in any of the yields.

AKNOWLEDGEMENTS

This work was funded by the Robert A. Welch Foundation Grant No. A1546, the NSF Grants No. 1058510 and No. 0722800, and the Qatar Foundation under the grant NPRP 09-585-1-087.

CAPTIONS

FIG. 1. (Color online) Typical TOF spectrum of methane irradiated with 50 fs, 800 nm pulses at a peak laser intensity of $\sim 6.3 \times 10^{14}$ W/cm$^2$. The different species are labeled. The data was taken with the polarization parallel to the TOF axis. Fragment clustering indicates two main ionization stages $CH_n^+$ and $CH_n^{2+}$.

FIG. 2. (Color online) (a) Intensity-resolved ionization yields of methane. The data consist of the parent ions and their daughter fragments. For each datum, the ionic species is denoted by the legend. This data was taken with the laser polarization perpendicular to the TOF axis. Consecutive shifts of the fragments toward higher intensities indicate that dissociation of $CH_n^+$ follows a step-wise process $CH_n^+ \rightarrow CH_{n-1}^+ + H$. (b) Calculated ionization probabilities using ADK and PPT theories. The theoretical ionizations probabilities for $CH_4^{2+}$ have been scaled by 0.04 for comparison with the measured data.

FIG. 3. (Color online) Angular dependent ionization yields of $CH_n^+$ [$n = 1-3$, panel (a—c)]. All yields have been normalized to $CH_4^+$ to compensate for pressure effects. The data was taken for a peak laser intensity of $\sim 1.2 \times 10^{13}$ W/cm$^2$. For this intensity and statistics all yields are isotropic.

FIG. 4. (Color online). Angular-dependent ion yields of $CH_n^+$ [$n = 4-1$, panel (a—d)] and $CH_{2n}^{2+}$ [$n = 1,0$, panel (e, f)] taken at an intensity of $\sim 3.6 \times 10^{14}$ W/cm$^2$ and normalized to $CH_4^+$ to compensate for pressure effects. All fragment ion yields exhibit an anisotropic behavior. The yields have been fit to $r = r_0 + a\cos^{2n}(\theta) + b\sin^{2m}(\theta)$ [black curve], where $r_0$ (red curve), $a$ (blue curve), $b$ (green curve), $n$ and $m$ are fit parameters.

FIG. 5. (Color online). Angular-dependent ion yields of $CH_n^+$ [$n=4-1$, panel (a—d)] and $CH_{2n}^{2+}$ [$n=1,0$, panel (e, f)] taken at an intensity of $\sim 6.3\times10^{14}$ W/cm$^2$ and normalized to $CH_4^+$ to compensate for pressure effects. Fragmental $CH_2^{2+}$ (panel e) exhibits an anomalous perpendicular fragmentation pattern. The yields have been fit to $r = r_0 + a\cos^{2n}(\theta) + b\sin^{2m}(\theta)$, where $r_0$ (red curve), $a$ (blue curve), $b$ (green curve), $n$ and $m$ are fit parameters.

FIG. 6. (Color online). Images of the measured ions taken across the focus and as a function of polarization angle. Panels (a—e) are the singly-charged species $CH_n^+$ ($n=0-4$), panels (f—h) are the doubly-charged species $CH_{2n}^{2+}$ ($n=0-2$), and panel (i) is $H^+$. Each ionic species has been scaled to the position of hydrogen and then centered on zero. The highest intensities are therefore around zero. The anisotropies can be seen for all ionic species except for the parent ions. Panels (a—c, f) show a depletion at the higher intensities due to higher-order processes. Because of the high extraction voltage, the yields remain intact. Hydrogen shows shifting to later times when the polarization is parallel to the TOF axis.

FIG. 7. (Color online). Effects of kinetic energy release in the measured ions. Each ionic species has been fitted to a suitable function to track the change in arrival times. For comparison, these arrival times for the different species have been scaled to the expected position of hydrogen, found by calibrating the time scale to mass over charge. Using Eq. 1 the kinetic energy of $H^+$ was 39.7 eV (27 eV) when the laser polarization was parallel (perpendicular) to the TOF axis. For $H_2^+$ these values are 18.9 eV and 4.4 eV respectively.

Figure 1

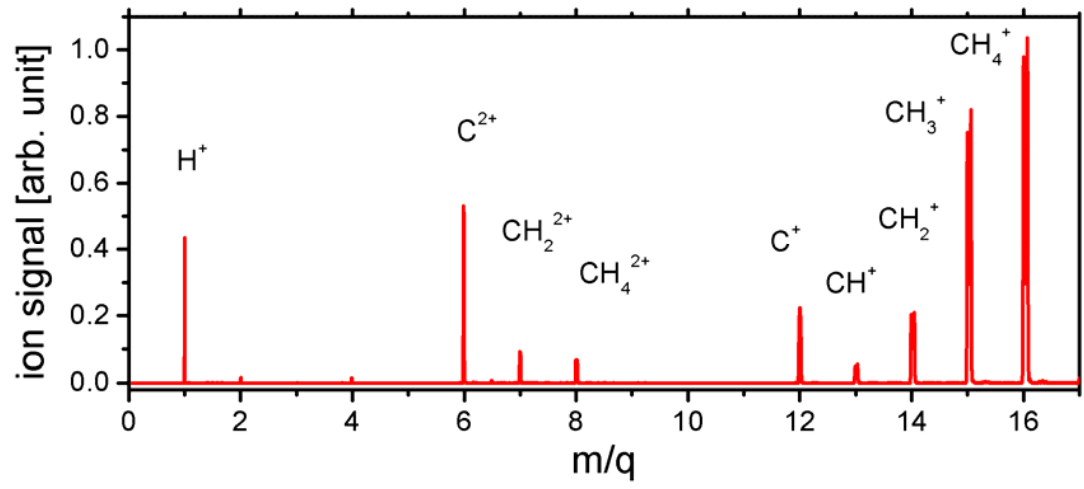

Figure 2

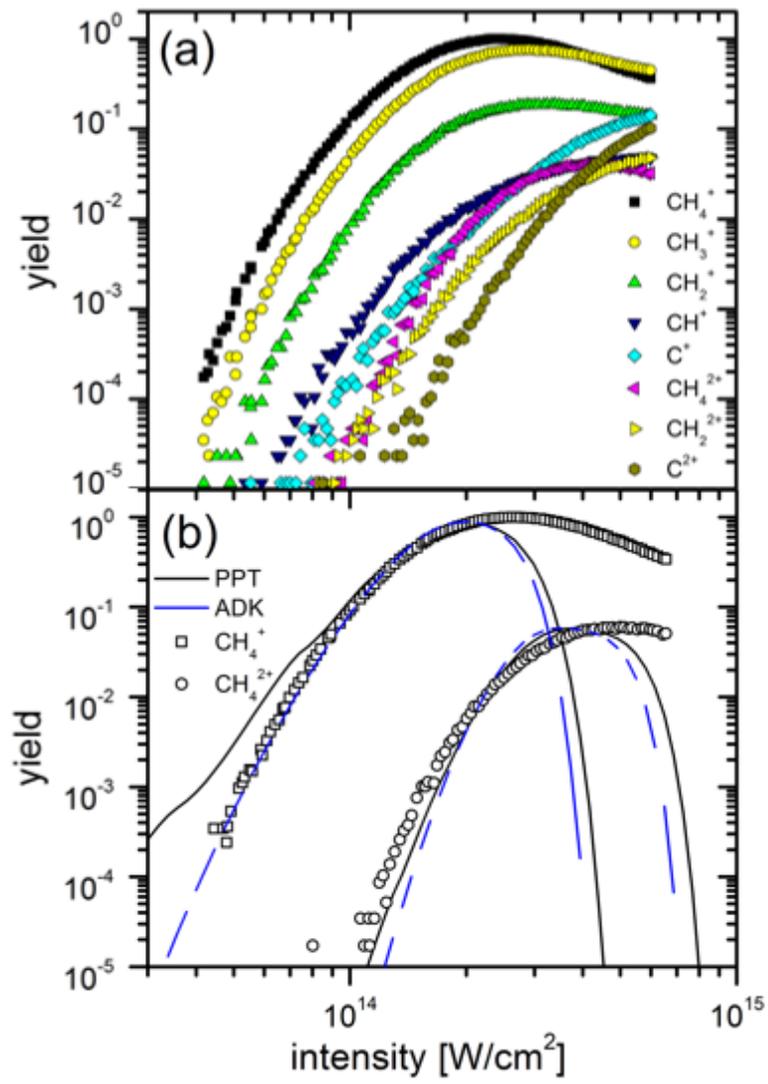

Figure 3

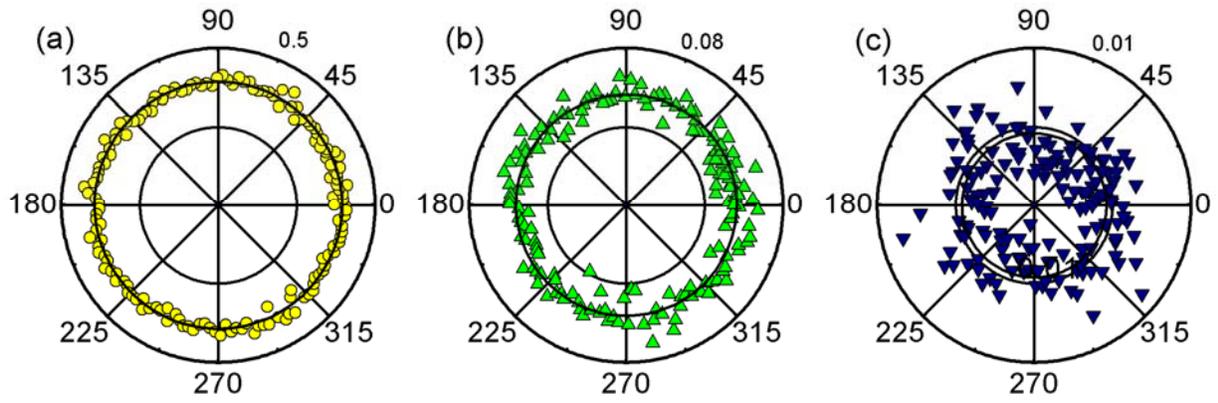

Figure 4

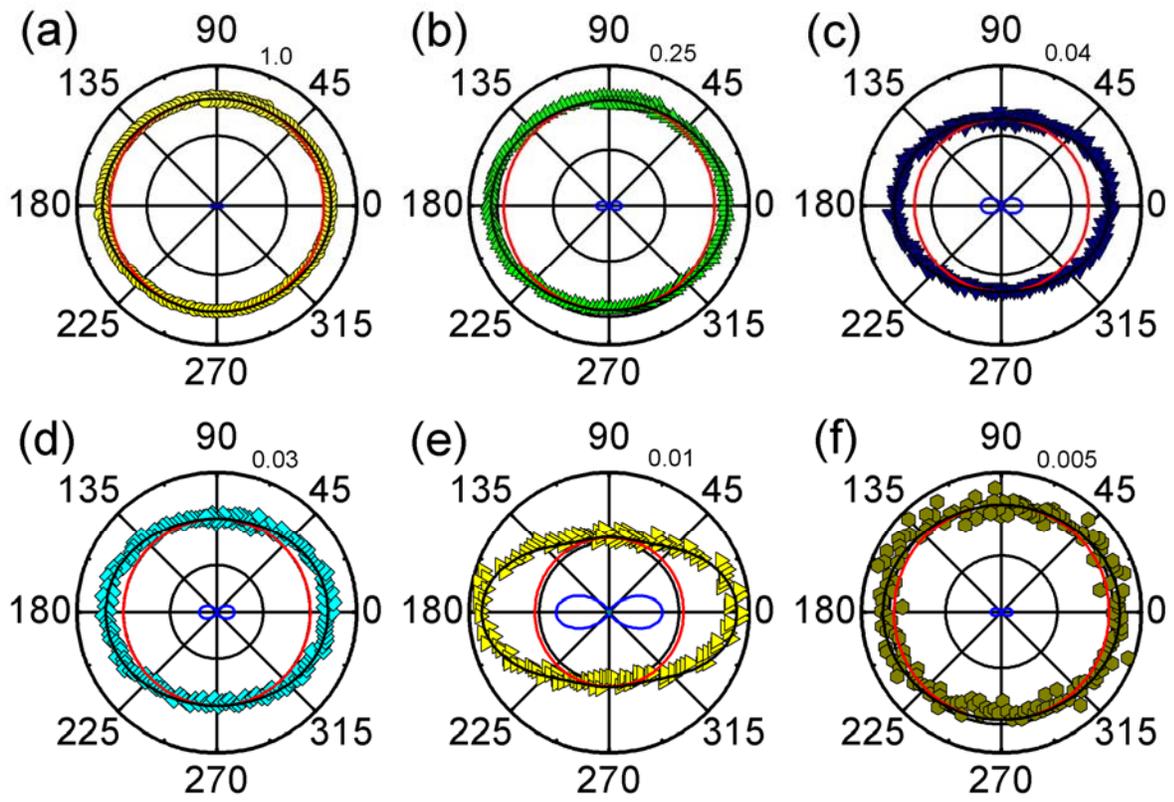

Figure 5

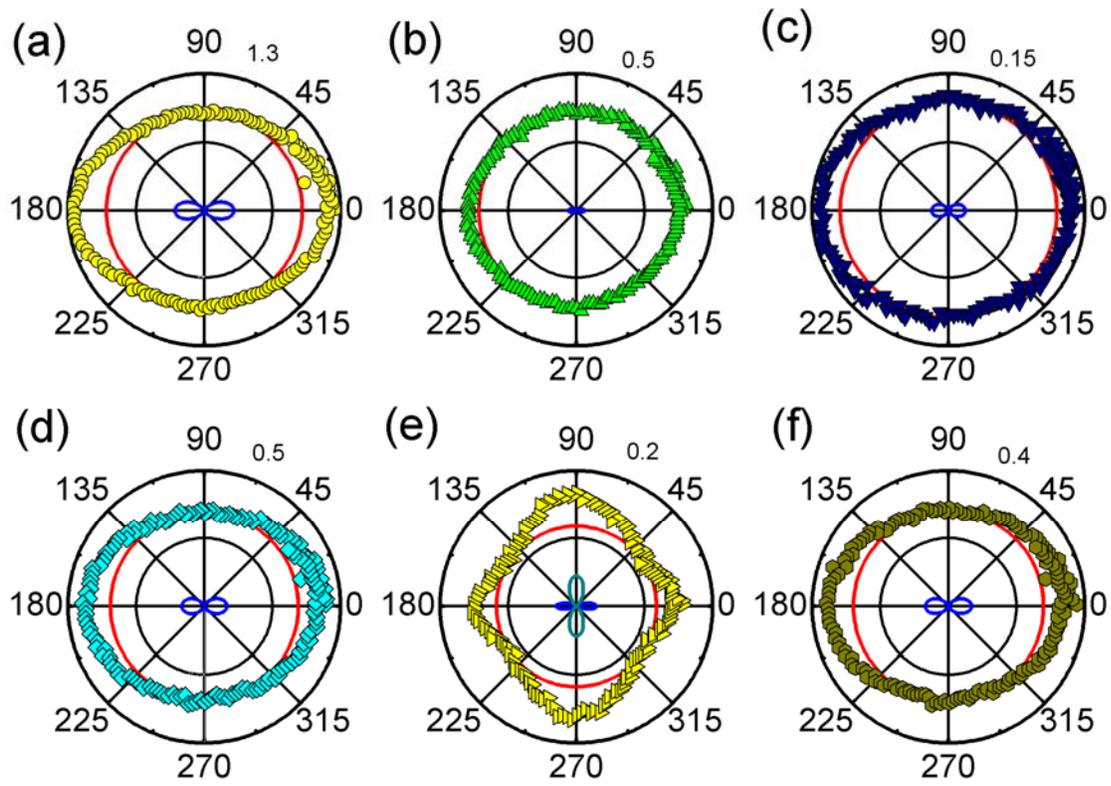

Fig. 6

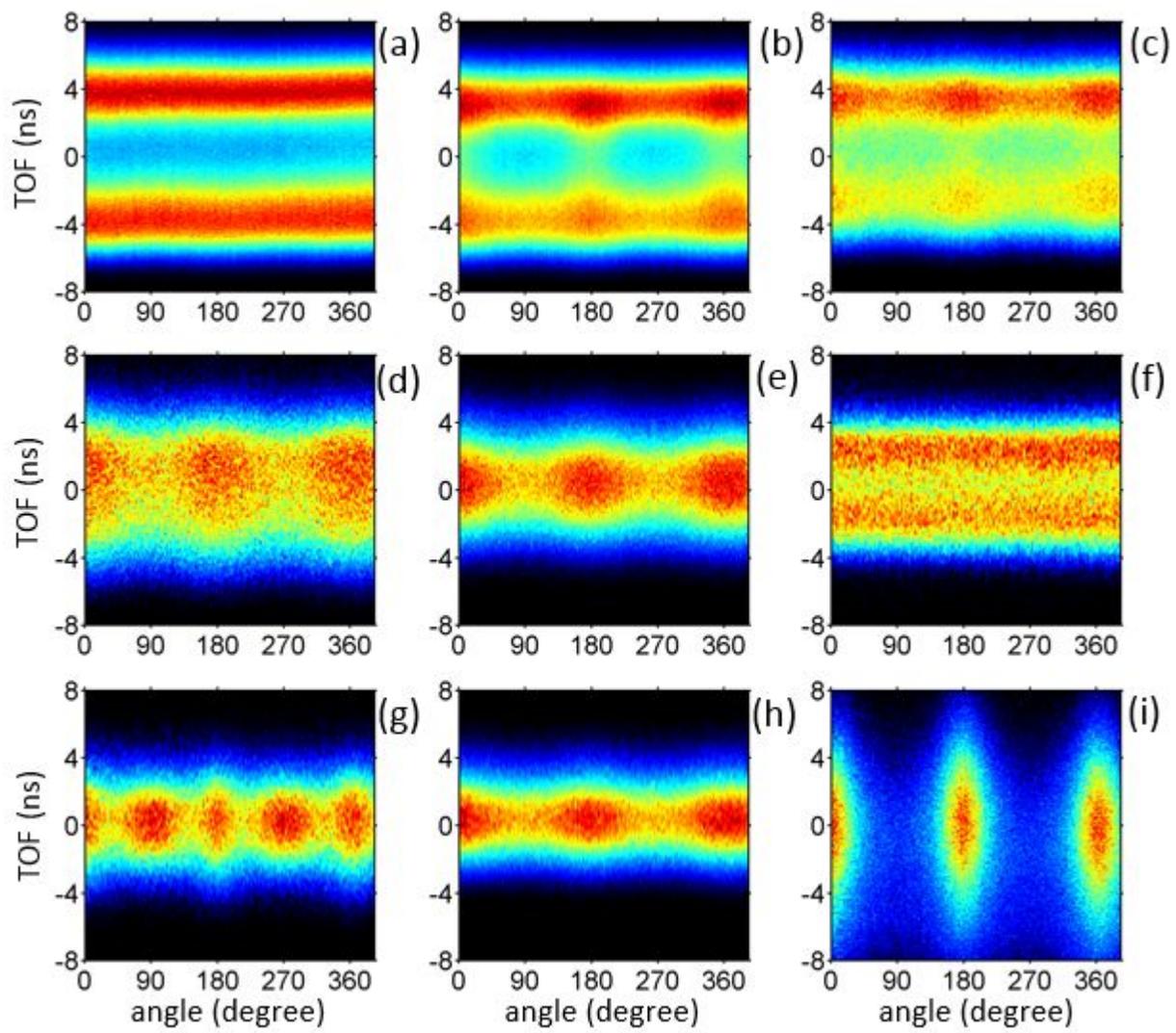

Figure 7

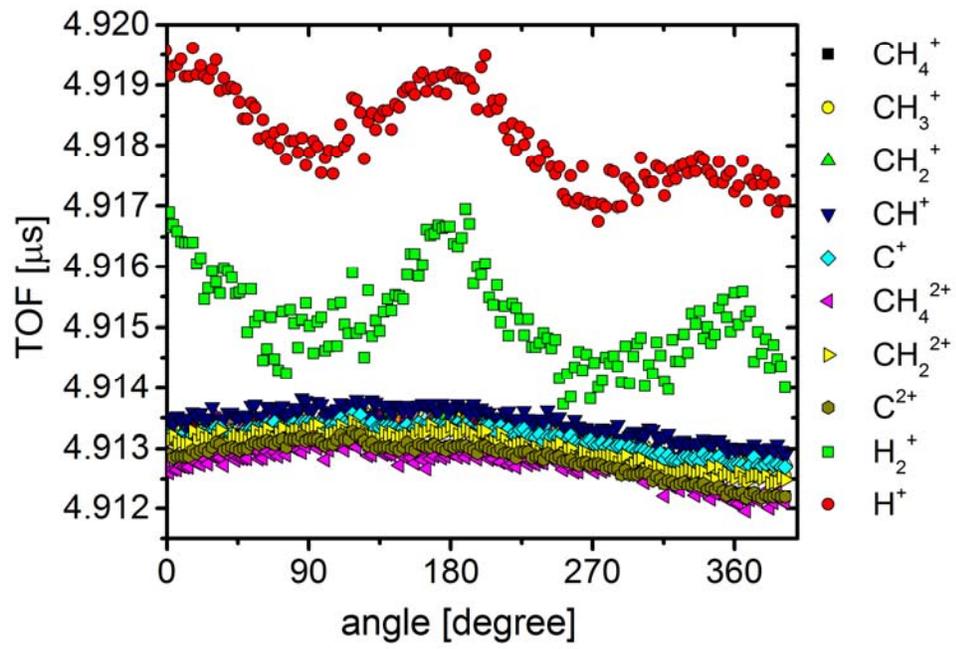